\documentclass[aps,pre,twocolumn,longbibliography]{revtex4-1}

\usepackage{bm}
\usepackage{graphicx}
\usepackage{amsmath}
\usepackage{amssymb}
\usepackage{amsfonts}
\usepackage{float}
\usepackage{color}
\usepackage[utf8x]{inputenc}

\begin{document}

\title{Information capacity of a network of spiking neurons.}

\author{Silvia Scarpetta$^{1,2}$}
\author{Antonio de Candia$^{2,3}$}

\affiliation{$^1$Dipartimento di Fisica ``E. Caianiello'', Universit\`a di Salerno, Fisciano (SA), Italy}
\affiliation{$^2$INFN, Sezione di Napoli, Gruppo collegato di Salerno, Fisciano (SA), Italy}
\affiliation{$^3$Dipartimento di Fisica ``E. Pancini'', Universit\`a di Napoli Federico II,\\
Complesso Universitario di Monte Sant'Angelo, via Cintia, 80126 Napoli, Italy}

\begin{abstract}
We study a model of spiking neurons, with recurrent connections that result from learning a set of spatio-temporal patterns
with a spike-timing dependent plasticity rule and a global inhibition.
We investigate the ability of the network to store and selectively replay multiple patterns of spikes,
with a combination of spatial population and phase-of-spike code.
Each neuron in a pattern is characterized by a binary variable determining
if the neuron is active in the pattern, and a phase-lag variable representing the spike-timing order among the active units.
After the learning stage, we study the dynamics of the network induced by a brief cue stimulation,
and verify that the network is able to selectively replay the pattern correctly and persistently.
We calculate the information capacity of the network, defined as the maximum number of patterns that can
be encoded in the network times the number of bits carried
by each pattern, normalized by the number of synapses, and find that it can reach a value $\alpha_\text{max}\simeq 0.27$, similar to the one of 
sequence processing neural networks, and almost double of the capacity of the static Hopfield model.
We study the dependence of the capacity on the global inhibition, connection strength
(or neuron threshold)
and fraction of neurons participating to the patterns. The results show that a dual population and temporal coding can be optimal for the capacity of an associative memory.
\end{abstract}

\maketitle

\section{Introduction}
 
Precise spatio-temporal patterns  of spikes, where information is coded both in the population (spatial) distribution and in the
 precise timing of the spikes,
has been hypothesized to play a fundamental role in information coding, processing and memory in the
 brain \cite{ref1,ref5, ref5z,ref6}.

  Notably, precise spatio-temporal  memory traces, stored during experience,
  are replayed during post-experience sleep,
  with a different time scale but with the same phase-relationship  \cite{ref2,Wilson,Mehta},
It has been conjectured that spontaneous replay of stored patterns during sleep is relevant for memory consolidation.

Indeed, an important feature of cortical activity, reported in a variety of in
 vivo studies, is the similarity of spontaneous and sensory-evoked activity patterns
 \cite{ref14a,ref15a,ref16a}
 Recently the spontaneous and evoked activity similarity has been reliably
 observed also in dissociated cortical networks  \cite{ref17a}, reinforcing the idea that the
 emergence of cortical recurring patterns, both during spontaneous and evoked activity,
 is the result of the cortical connectivity (with its micro-circuits and dynamical
 attractors) \cite{ref17a}.

 Several neural codes (firing rate, population code, temporal code, phase of spike code, and combinations of these)
 have been proposed in order to explain how neurons encode sensory information.
 For long time the rate code hypothesis has been considered adequate,
 but there is increasing evidence that
 the timing of spikes relative to others emitted by the same or other neurons may be highly relevant in many situations
 \cite{ref5z,ref6,ref5c,mazzoni,huxternature}.
 
Recently \cite{ref5z},
the performance of different neural codes in encoding natural sounds in the auditory cortex of alert primates has been compared.
 It has been  demonstrated that both temporal spike patterns and spatial population codes provide more information than firing rates alone,
 and that they provide complementary informations.
 Results  in primary visual and auditory cortices show that period of slow
oscillatory network activity
can be used as internal reference frame to effectively partition spike trains and extract phase information
\cite{ref5b}.
 Combining both spatial population activity with
 temporal coding  mechanisms
 into a dual code
provides significant gains of information \cite{ref5z,mazzoni,huxternature}.
A dual code consisting of adding a phase label to population code was found to be not only more informative,
but also more robust to sensory noise  \cite{ref5z}.

Following these experimental results,
in this paper we study storage and retrieval of spatio-temporal patterns of activity, where
each pattern is formed by a subset of the neurons, firing in a precise temporal order.
Therefore, in each pattern the information is coded both in the binary spatial part, indicating
   which units are active (as in the Hopfield model), and in the temporal part,
   indicating the spiking order and the precise phase relationships of the units that are active in that pattern.

A fundamental ingredient of the model is the spike-timing dependent plasticity (STDP) rule that shapes the connections between neurons.
In the first learning stage, the network is forced to replay the patterns to be encoded, and the STDP learning rule is applied to determine the connections.
While within the class of usual autoassociative Hebbian rules (as in the Hopfield model) the connections are symmetric, and therefore the
dynamics of the network relaxes towards static patterns, in our case the plasticity
crucially depends on the temporal order of the spikings, and results in asymmetric connections between neurons, so that the attractors of the
dynamics are spatio-temporal patterns with a defined order of spiking of the neurons.
We add, to the connection determined by the STDP rule, a term representing a global inhibition, that reduces the probability that
a neuron not participating to the pattern fires
incorrectly. 

The model is different from other models of memory storage of dynamical patterns, as sequence processing neural networks \cite{During,Leibold}
and synfire chains \cite{Synfire1,Synfire2,Synfire3,Synfire4}, because in 
such models patterns are characterized by pools of neurons that fire synchronously, each pool can belong only to one chain,
and each neuron can belong to more than one pool, thus
firing at different times within the pattern.
 In our case, a pattern is characterized by a pool of neurons that are inactive during all pattern evolution,
  while other neurons fire sequentially,
   so that no synchrony between neurons is required,
and each active neuron is characterized by a precise time of firing with respect to other active neurons within the pattern.

Therefore our
model is more similar to the Polychronization model
introduced by Izhichevich \cite{Poly}  where precise reproducible
time-locking patterns emerges in network dynamics.
Differently from our model, in the model of Izhichevich
polychronous groups are not stable attractors from dynamical
system point of view, they emerge with a stereotypical but
transient activity that typically lasts less then 100 ms \cite{Poly}.

\section{The model} 

We simulate a network of $N$ leaky integrate-and-fire neurons, where a number $P$ of spatio-temporal pattern are encoded in the connections.
The activity of the neuron $j=1,\ldots,N$ in pattern $\mu=1,\ldots,P$ is given by the periodic spike train
 \[
x_j^\mu(t)= \xi_j^\mu\sum\limits_{n=-\infty}^\infty\delta\!\left[t-  \left(\frac{\phi_j^\mu}{2\pi} + n\right) T^\mu\right],
 \]
 where $\xi_j^\mu=0,1$
 are randomly and independently drawn binary variables indicating if neuron $j$ participates or not to the pattern,
 $0\le \phi_j^\mu<2\pi$ 
 are randomly and independently drawn variables indicating the phases of firing,
 $T^\mu $ is the period of the pattern,
 and the integer number $n$ spans all the spikes emitted by the neuron.
We extract randomly $P$ patterns, where $0<M\le N$ of the neurons are active, that is have $\xi_j^\mu=1$, while
the remaining $N-M$ have $\xi_j^\mu=0$, and we extract randomly the phases $\phi_j^\mu$ for the $M$ active neurons.

\begin{figure}[h]
\begin{center}
\includegraphics[width=6cm]{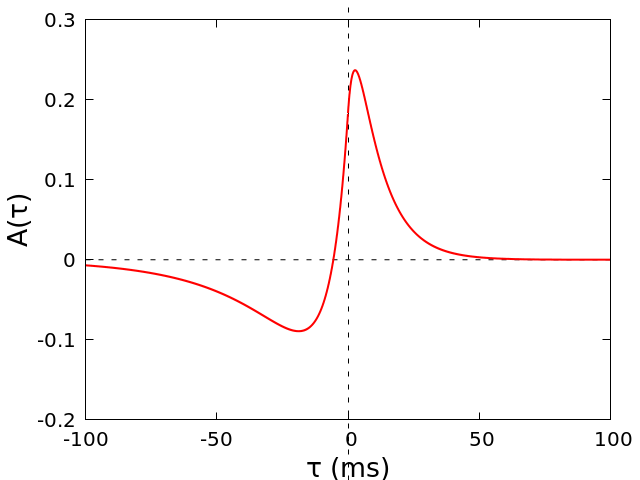}
\end{center}
\caption{The learning kernel $A(\tau)=a_pe^{-\tau/t_p}-a_de^{-\eta\tau/t_p}$ if $\tau>0$, and $A(\tau)=a_pe^{\eta\tau/t_d}-a_de^{\tau/t_d}$ if $\tau<0$,
with $a_p=(1+\eta t_p/t_d)^{-1}$, $a_d=(\eta+t_p/t_d)^{-1}$, $\eta=4$, $t_p=10.2\>\text{ms}$, $t_d=28.6\>\text{ms}$. The parameters are the same used in
\cite{abarbanel} to fit the experimental data of  \cite{ref23a}.}
\label{figkernel}
\end{figure}

Then, in the first learning stage, we set the connection between
pre-synaptic neuron $i$ and post-synaptic neuron $j$ using  the learning rule
\begin{equation}
J_{ij}=-I_0 + E_0\sum_{\mu=1}^P\frac{T^\mu}{\hat{T}}\int\limits_0^{\hat{T}}\!dt\!\int\limits_0^{\hat{T}}\!dt^\prime\, x_i^\mu(t^\prime)A(t-t^\prime)x_j^\mu(t),
\label{learningrule}
\end{equation}
where $\hat{T}\gg T^\mu$ is the learning time, 
$I_0$ is an overall inhibition, $E_0$ is a parameter determining the strength of the connections for co-active neurons,
and $A(\tau)$ is a spike-timing dependent plasticity (STDP) kernel, that
makes the connection dependent on the precise relative timing of the pre-synaptic and post-synaptic activity,
as observed in neocortical and hippocampal pyramidal cells \cite{ref22a,ref23a,ref26a}.
As shown in Fig.\ \ref{figkernel}, 
the kernel is such that the connection between $i$ (pre) and $j$ (post) is potentiated if neuron $j$
fires few milliseconds after neuron $i$, and depressed if the order is reversed. Note that the normalization $T^\mu/\hat{T}$ ensures that the connections do not depend
on the learning time $\hat{T}$, if it is large enough that border effect can be neglected.
This rule extends the rule used in \cite{noiPLOS2013,noi2018}, with the addition of the global inhibition term $I_0$, that inhibits the activity of neurons not participating
to the pattern.

After learning a number $P$ of patterns with the learning rule (\ref{learningrule}), we simulate the dynamics of the
network by considering a leaky integrate-and-fire model of neurons, where the membrane potential obeys the
equation
\[
\frac{dV_j}{dt}=-V_j(t)/\tau_m+I_j(t)
\]
(membrane capacity is set to one), with
\[
I_j(t)=\sum_{\hat{t}_i>\hat{t}_j} J_{ij} e^{-(t-\hat{t}_i)/\tau_s},
\]
where $\hat{t}_i>\hat{t}_j$ are all the spikes in input to the neuron $j$ after the last spike of neuron $j$,
membrane time constant $\tau_m=10\>\text{ms}$ and synapse time constant $\tau_s=5\>\text{ms}$.
When the potential reaches the threshold $\Theta$ (conventionally set to $\Theta=1$), the neuron fires and the
potential is reset to the resting value $V=0$.
Note that, among the parameters $I_0$, $E_0$ and $\Theta$, only two are independent. We have chosen to set $\Theta=1$, but one could as well set for example $E_0=1$, and
study the dependence of the model with respect to $I_0$ and $\Theta$.

The dynamics of the network is started by a short cue, i.e. a short train of $H\ll M$ spikes forced in a subset of neurons
with the order given by one of the stored patterns. The times of this train of spikes are given by $t_{i,\text{cue}}=(i/N)T_{\text{cue}}$, with $i=1\ldots H$,
where $T_{\text{cue}}$ can be different from the period $T^\mu$ of the pattern.
In the following the number of neurons is $N=6000$, if not otherwise stated,
the period of the patterns is $T^\mu=125\>\text{ms}$, $H=M/10$, and $T_{\text{cue}}=83\>\text{ms}$.

\section{Results}

\begin{figure}[h]
\begin{center}
\includegraphics[width=6cm]{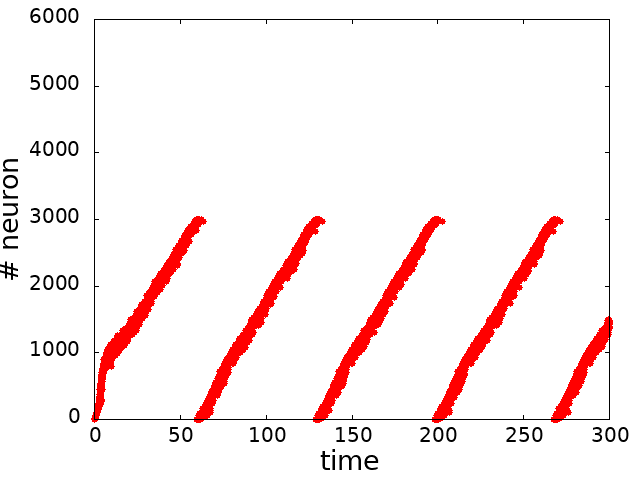}
\includegraphics[width=6cm]{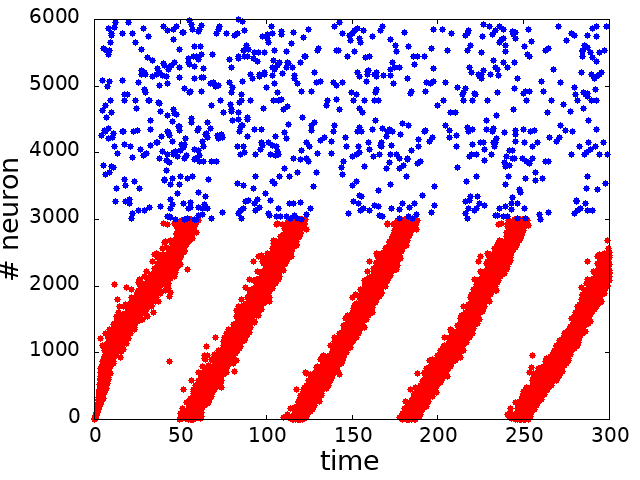}
\end{center}
\caption{Two examples of successful replay, initiated by a stimulus cue, for the network with $N=6000$ and $M=3000$.
Each dot represents a spike of the neuron indicated by the y-axis at the time indicated by the x-axis.
On the y-axis the first 3000 neurons are the ones participating to the pattern being replayed, with increasing phase of firing, and the following 3000
are the neurons not participating to the pattern. In both cases A and B, the parameters used are $I_0=0.0133$ and $E_0=0.2856$,
the length of the cue is $H=300$ and its period $T_{\text{cue}}=83\>\text{ms}$.
A) In this case the number of patterns is $P=30$, and the order parameter is $m=0.995$. None of the neurons not participating to the pattern fires
during the dynamics.
B) The number of patterns is $P=180$, and the order parameter is $m=0.938$. In this case there are several spikes also from the neurons that do not
participate to the pattern, as an effect of the worse quality of the replay.}
\label{figspike}
\end{figure}

After inducing the short cue in the network, we observe the subsequent spontaneous dynamics, to see if the network is
able to reproduce completely and persistently the pattern being stimulated by the cue.
In Fig.\ \ref{figspike}, we show two examples of successful replay with $M=3000$, that is half of the neurons are active in each pattern.
In the case of Fig.\ \ref{figspike}A, the number of stored patterns is $P=30$.
On the y-axis the first 3000 neurons are the ones participating to the pattern, with increasing phase of firing, and the following 3000
are the ones not participating.
The raster plot shows that, after the cue that lasts $\sim$ 4 ms, the neurons continue to fire
approximately in the order corresponding to the pattern that has been stimulated, that therefore has been
retrieved with high fidelity. On the other hand, in Fig.\ \ref{figspike}B the number of stored patterns is $P=180$, and the quality of the replay
is worse. The order of firing of the neurons is less precise, as shown by the thickening of the plot, and several neurons fire although they do not belong
to the pattern being retrieved. With these parameters, when the number of stored patterns becomes larger than $P_{\text{max}}\simeq 200$,
the network is not able to retrieve any of the patterns, and the dynamics following the cue becomes completely chaotic.

In order to quantify the quality of the selective cue-induced replay
of the pattern, we define the overlap $q^\mu$ as
\begin{equation}
q^\mu([t_0,t_1])=\max_{T^w}\left|\frac{1}{N_s}\sum_{j=1}^M\sum_{t_0\le t_j^\ast\le t_1}e^{2\pi it_j^\ast/T^w}e^{-2\pi it_j^\mu/T^\mu}\right|,
\label{qequ}
\end{equation}
where the sum is done over the spikes $t_j^\ast$ emitted  in a given interval of time $[t_0,t_1]$ by the neurons belonging to the pattern $\mu$,
$N_s$ is the total number of spikes (included those emitted by neurons not belonging to the pattern)
and we take the maximum over the time window $T^w$. The overlap tends to the value $q^\mu=1$ if the neurons belonging to the pattern fire
with the same order of the pattern, with a period of replay $T^w$ in general different from the period $T^\mu$ characterizing the pattern in the learning stage,
and those not belonging to the pattern do not fire.
When the quality of the replay is worse, because some of the neurons not belonging to the pattern fire, or if the order of replay is not exact,
the value of the overlap $q^\mu$ is lower than one.
Note that the order parameter does not make any assumption on the time scale of the replay,
in agreement with experimental results showing that
the reactivation of memory traces from past experience  happens
on a compressed timescale during sleep and awake state  \cite{ref5,ref2,Wilson,Mehta}.

We compute the overlap in the interval $[t_0,t_1]=[100\>\text{ms},300\>\text{ms}]$.
In the case of Fig.\ \ref{figspike}, the value of the overlap is $q^\mu=0.995$ in Fig.\ \ref{figspike}A
and $q^\mu=0.938$ in Fig.\ \ref{figspike}B.

\begin{figure}[h]
\begin{center}
\includegraphics[width=6cm]{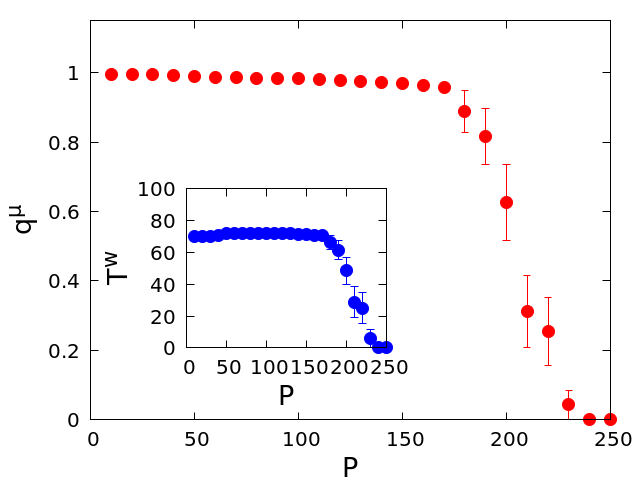}
\end{center}
\caption{Plot of the overlap $q^\mu$ as a function of the number $P$ of patterns encoded in the network, for $N=6000$, $M=3000$,
$I_0=0.0133$, $E_0=0.2856$, $H=300$, $T_{\text{cue}}=83\>\text{ms}$. Inset: period of the replay $T^w$.}
\label{figoverlap}
\end{figure}

In Fig.\ \ref{figoverlap} we show the value of the overlap as a function of the number of stored patterns.
The overlap abruptly goes from a value close to one, when the replay is successful, to a value close to
zero when replay fails. Failure may happen both because there is not sustained activity at all,
 or because the sustained activity is chaotic and not similar to the pattern to be retrieved.
We can measure the storage capacity of the network as the maximum number $P_{\text{max}}$ of patterns that can be stored and retrieved with an overlap
greater or equal to 0.5. The storage capacity for the parameters of Fig.\ \ref{figoverlap} therefore is $P_{\text{max}}=200$.
In the Inset, we show the period of replay $T^w$, that is approximately constant for almost all values of the number $P$ of patterns encoded, and
decreases to zero when $P$ gets near to its maximum value.

We have then studied the capacity of the network as a function of the network parameters $I_0$ and $E_0$, for different number  $M$ of active neurons.
To compare the capacity for different values of $M$ (and with the case of static patterns, for example with the Hopfield model),
it is convenient to consider the information content of each pattern encoded, that is the number of bits carried by each pattern.
This is given by the base two logarithm of the
number of different patterns that can be encoded. For the Hopfield model we have $2^N$ possible patterns, so that the number of bits of each pattern is
$B=\log_2 2^N=N$.

In the case of dynamical patterns considered here, where $M$ of the $N$ neurons are active, we have $\binom{N}{M}$ possible choices of the active neurons, and $M!$ possible
sequences of the active neurons, so that the number of bits carried by a single pattern is given by
\[
B=\log_2\left[\binom{N}{M}M!\right]\simeq M\log_2 N.
\]

We therefore define the information load per synapse of the network as $\alpha=PB/N^2$, where $P$ is the number of patterns encoded and $B$ the number of bits of each patterns.
Note that
in the case of the Hopfield model this gives the usual definition $\alpha=P/N$.
The information capacity is defined as the maximum value of the load, $\alpha_{\text{max}}=P_{\text{max}}B/N^2$,
such that the patterns can be effectively retrieved.

\begin{figure}[h]
\begin{center}
\includegraphics[width=6cm]{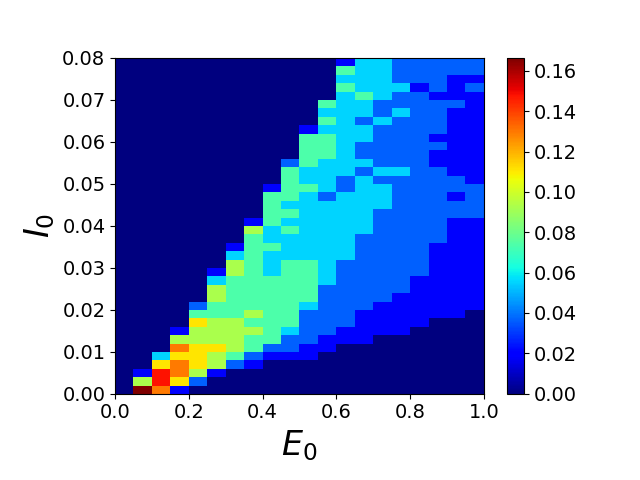}
\includegraphics[width=6cm]{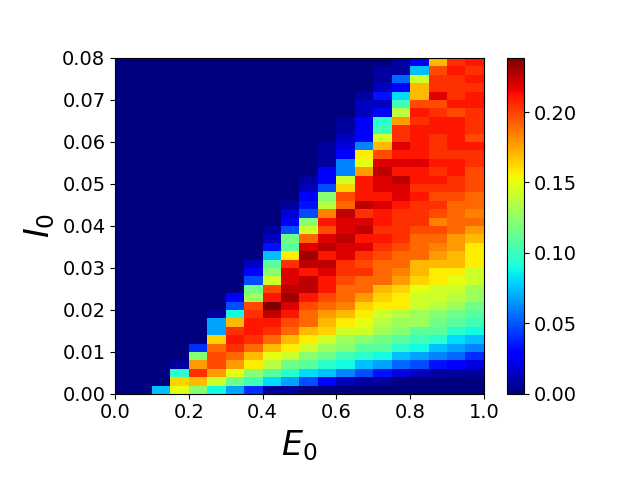}
\includegraphics[width=6cm]{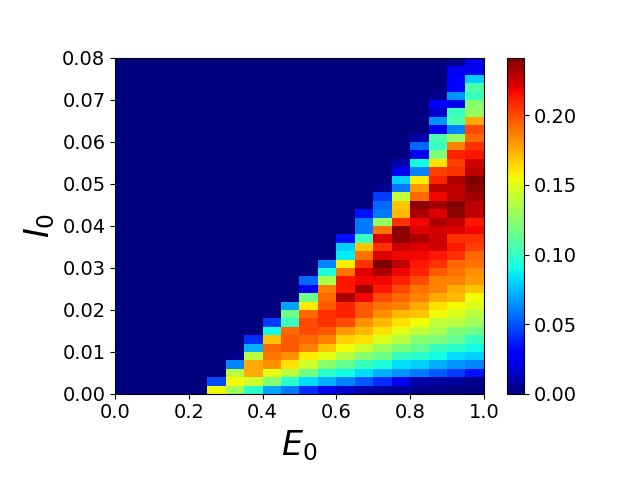}
\end{center}
\caption{Plot of the information capacity $\alpha_{\text{max}}=P_{\text{max}}B/N^2$ of a network of $N=6000$ neurons, as a function of the overall inhibition $I_0$
and connection strength $E_0$, for A) $M=6000$, B) $M=2000$ and C) $M=1000$.}
\label{figpalette}
\end{figure}

In Fig.\ \ref{figpalette}, we show the information capacity $\alpha_{\text{max}}=P_{\text{max}}B/N^2$
of a network with $N=6000$ neurons, as a function of the overall inhibition $I_0$, and connection strength $E_0$,
for A) $M=6000$, B) $M=2000$ and C) $M=1000$, with a cue of length $H=M/10$ spikes.
In the case $M=6000$, when all the neurons participate to the pattern, the maximum capacity is $\alpha_{\text{max}}=0.167$, that is obtained
for a value $I_0\simeq 0$ of the global inhibition, and $E_0\simeq 0.1$.

When the number $M$ of neurons in the pattern becomes lower than $N$,
the maximum of the capacity is obtained for increasing values of both $I_0$ and $E_0$. For $M=2000$ we find a maximum $\alpha_{\text{max}}=0.239$ at
$I_0\simeq 0.022$ and $E_0\simeq 0.45$, while for $M=1000$ the maximum is $\alpha_{\text{max}}=0.248$ at
$I_0\simeq 0.054$ and $E_0\simeq 0.95$.
The main reason for which higher values of $E_0$ are needed for lower $M$, is to compensate the lower number of terms of the sum in Eq.\ (\ref{learningrule}),
at fixed neuron threshold.
On the other hand, the global inhibition serves to avoid the firing of neurons that do not participate to the pattern being replayed,
and therefore increases with he ratio $(N-M)/M$ between the neurons participating or not.

\begin{figure}[h]
\begin{center}
\includegraphics[width=6cm]{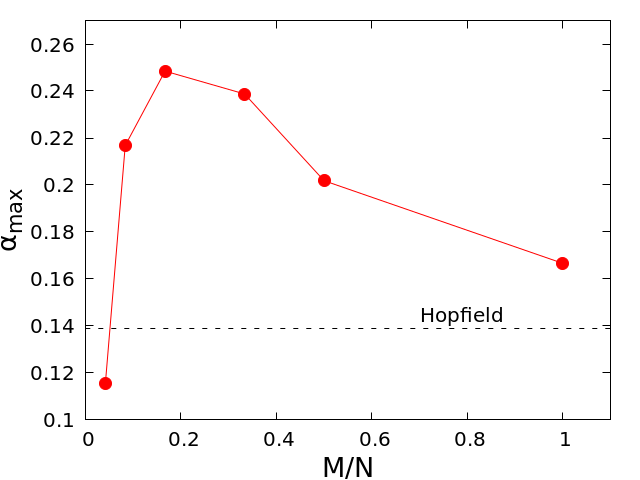}
\end{center}
\caption{Maximum value of the information capacity as a function of the ratio $M/N$, where $N=6000$ is the total number of neurons of the network, and $M$ the number of neurons
participating to the patterns.}
\label{figmax}
\end{figure}

In Fig.\ \ref{figmax}, we show the maximum value of the capacity as a function of the fraction $M/N$.
The overall maximum capacities are obtained for values of 
the fraction of neurons participating to the patterns around $M/N\simeq 0.2$.
This means  that storage of dual population and phase coding patterns is more efficient than
phase coding alone ($M/N=1$, where the pattern is defined only by the phases $\phi_j^\mu$ \cite{noi2010,noi2011,noi2013}). Notably the
information capacity of such dual coding is enhanced also
with respect to the binary population coding alone of the Hopfield model (where the pattern is defined only by the activities $\xi_j^\mu$).

\begin{figure}[h]
\begin{center}
\includegraphics[width=6cm]{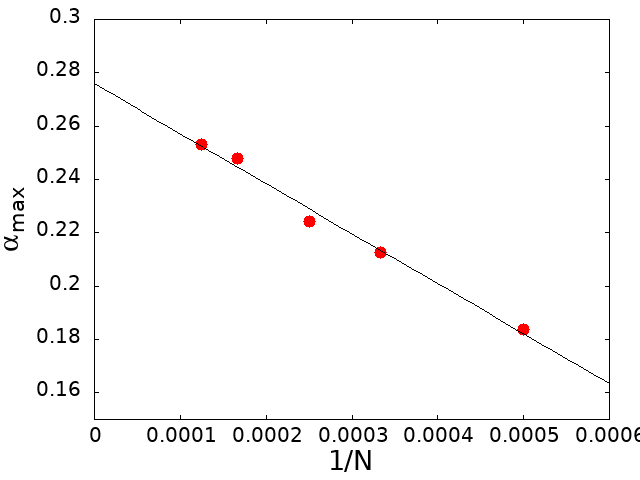}
\end{center}
\caption{Dependence of the maximum value of the  information capacity from the size of the network, for a fraction $M/N=0.2$ of active neurons.
The best fit is given by $\alpha_{\text{max}}=0.276-187/N$.}
\label{figsize}
\end{figure}

In Fig.\ \ref{figsize}, we show the maximum capacity as a function of the network size $N$, for a fixed value $M/N=0.2$ of the fraction of neurons participating to the pattern.
The capacity grows with the size of the network, and tends to a value $\alpha_{\text{max}}=0.276$ for very large sizes.
Notably the capacity is almost  the double of that of the Hopfield model ($\alpha_{\text{max}}=0.139$), and
practically equivalent (slightly larger) than that of the binary sequence processing Hopfield-like  model
($\alpha_{\text{max}}=0.269$) \cite{During}.

\section{Discussion}
In this paper we have studied the information capacity of a spiking network, able to store and selectively retrieve patterns defined
by a binary spatial population coding, as in the Hopfield
model, nested with a phase-of-firing coding.
The spatial population binary variable $\xi_j^\mu$ determine
the $M$ active neurons in pattern $\mu$, while the phase-of-firing
variable $\phi_j^\mu$ define the precise relative timing of firing of those active neurons.
The success of retrieval is measured by the order parameter $q^\mu$, introduced in eq. (\ref{qequ}), that extends the order
parameter of Hopfield model, since it requires both that
(1) the $N-M$ neurons silent in the pattern are silent in the retrieval,
and (2) the $M$ neurons active in the pattern  emits a spike
with the right phase relationship $\phi_j^\mu$ (with possibly a
different period of oscillation). The order parameter is
insensitive to the time scale of the replay, in agreement
with experimental results showing that the reactivation
of memory traces from past experience happens on a compressed timescale during sleep and awake state\cite{ref5,ref2,Wilson,Mehta}.

To study the memory capacity,
we measure the maximum number $P_{\text{max}}$
of stored patterns that can be successfully and selectively retrieved, when a cue is presented,  and
 in order to compare capacity of networks with different value of $M$, also
the maximum information capacity $\alpha_{\text{max}}$, i.e. the amount of information
(measured as the maximum number of patterns times the number of bits of each pattern) per synapse.

The ability to work as an associative memory for such spatio-temporal  patterns
is crucially dependent from the learning rule introduced in Eq.\ (\ref{learningrule}) where, in addition to
an overall inhibition, the connections between neurons co-active in a pattern are determined by a kernel depending on the precise timing of the spikes
between the pre-synaptic and post-synaptic neurons.
Note that this rule follows more closely the idea of Hebb, with respect to the Hopfield case, since connections between
neurons that are not active in the same pattern have
a negative weight contribution $-I_0$, while neurons which are both active in the same pattern
have a contribution that is positive if the pre-synaptic neuron fires few milliseconds before the post-synaptic one,
or negative if the reverse is true.
  The rule  in Eq.\ (\ref{learningrule}) extends the rule
  \cite{noi2002,noi2007,noi2010,noi2011,noi2013,noiPLOS2013,noi2018} used
  for storing  phase-coding-only patterns (i.e. $M/N=1$), and include a crucial global inhibition term, allowing the
  storage of patterns with dual coding $M/N<1$.

Notably here we found that a fraction $M/N\simeq 0.2$ gives the highest information capacity,
meaning that the storage of dual coding patterns, where information is stored both in the population (spatial) distribution of active neurons
and in the timing of the spikes, is more efficient then phase coding alone ($M/N=1$).
At $M/N\simeq 0.2$ not only the capacity is higher, but the region of the parameters $E_0$ and $I_0$ where capacity is near-maximal is larger,
so that a fine-tuning of the parameters is less important.

We also found that a crucial role is played by the value of the global inhibition $I_0$. When the fraction of neurons participating to the pattern
is smaller, higher values of $I_0$ are needed in order to reduce the occurrence of ``false positives'', that is neurons not participating to the
pattern that fire incorrectly.

We find that the maximum capacity at $M/N=0.2$ grows with $N$, and tends to $\alpha_{max}=0.276$ for very large size, a value close (slightly higher) than
the maximum capacity found for sequence processing networks, where discrete time dynamics, binary patterns sequence, with a Hopfield-like asymmetric
learning rule was studied.
It's interesting that there seems to be a unique maximum  information capacity for
storing dynamical patterns, in which time has been introduced in different ways,
the sequence of binary patterns with Hopfield-like chain learning rule $\xi^\mu \xi^{\mu+1}$ \cite{During},
and the dual coding patterns of spike defined here,
with a biologically inspired STDP-based learning rule.

These findings may help to elucidate experimental observations where both spatial and temporal coding schemes are found to be present,
representing different independent variables which are simulataneously encoded and binded together in the memorized patterns,
such as for example the location of the event and its behavioural and sensory content \cite{mazzoni,huxternature}.

\section{Acknowledgements}
A.d.C. acknowledges financial support of the MIUR PRIN 2017WZFTZP “Stochastic forecasting in complex systems”.

%\bibliographystyle{apsrev4-1}
%\bibliography{paper}{}

%merlin.mbs apsrev4-1.bst 2010-07-25 4.21a (PWD, AO, DPC) hacked
%Control: key (0)
%Control: author (72) initials jnrlst
%Control: editor formatted (1) identically to author
%Control: production of article title (-1) disabled
%Control: page (0) single
%Control: year (1) truncated
%Control: production of eprint (0) enabled
%

\end{document}